\documentclass[12pt]{article}

\usepackage{color}
\usepackage{times}
\usepackage{graphicx}
\usepackage{comment}
\usepackage[intlimits]{amsmath}
\usepackage{ulem}
\usepackage{bm}
\usepackage{xcolor}
\usepackage{epstopdf, epsfig}
\usepackage{amsmath}
\usepackage{amsthm}
\usepackage{textcomp}
\usepackage{amsfonts}
\usepackage{amssymb}
\usepackage{float}
\usepackage{soul}

\title{Rotation Control, Interlocking and Self-positioning of Active Cogwheels}

\author
{ Quentin Martinet,$^{1,2}$ Antoine Aubret,$^{1,3}$ Jeremie Palacci$^{1, 2\ast}$\\
\\
\normalsize{$^{1}$Department of Physics, University of California San Diego}\\
\normalsize{$^{2}$IST Austria, Am Campus 1, 3400 Klosterneuburg, Autria}\\
\normalsize{$^{3}$Current Address: Univ. Bordeaux, CNRS, LOMA, UMR 5798, F-33405 Talence, France}\\
}

\newcommand{\JP}[1]{\textcolor{black}{#1}}

\date{}

\begin{document} 

\baselineskip24pt

\maketitle

\begin{abstract} 
\textcolor{black}{
Gears and cogwheels restrain degrees of freedom and channel power into a specified motion. They are fundamental components of macroscopic machines. Interlocking microrotors similarly constitute key elements toward feasible micromachinery. Their assembly, positioning and control is a challenge at microscale, where noise is ubiquituous. Here, we show the assembly and control of a family of self-spinning cogwheels with varying teeth numbers and study interlocking mechanisms in systems of multiple cogwheels. The cogwheels are autonomous and active, with teeth formed by colloidal microswimmers that power the structure, and control its rotation rate. Leveraging the angular momentum of light with optical vortices, we control the direction of rotation of the cogwheels. We study pairs of interlocking cogwheels, that roll over each other in a random walk and curvature-dependent mobility. We leverage this feature to achieve self-positioning of cogwheels on structures with variable curvature and program microbots, notably demonstrating the ability to pick up, displace and release a load. This work highlights untapped opportunities of manufacturing at microscale using self-positioning components and constitutes an important step towards autonomous and programmable microbots.}
\end{abstract}
\section*{Introduction}
Biological nanomachines thrust life and dynamic processes across scales \cite{Houdusse_nanomotors}: Molecular motors organize chromosomes  \cite{chromosomes_loops}, play a key role in cell division and are responsible for muscular contraction \cite{Alberts:887198}. In contrast, our ability to realize  microstructures that rival their biological counterparts is still in its infancy. Challenges are manifold, requiring: {\it (i)} suitable fabrication and assembly strategies at small scale, {\it (ii)} development of tools of control and communication in noisy environment and {\it (iii)} power generation at micro and nanoscale. It calls for novel strategies marrying tools of basic science with the recent progress in robotics \cite{Li:2019fk,Rubenstein:2014dt}. The emerging field of active colloids -- microscale particles that transduce energy at small scale-- is uniquely suited to unlock those challenges and envision autonomous microbots of increasing complexity and versatility \cite{10.1021/jacs.1c04836, 10.1002/adma.202101965}. Progress in synthesis allows for bespoken design of microscale building blocks with programmable interactions \cite{He:2020kn, 10.1038/s41578-021-00323-x,Chaudhary:2012et,10.1038/natrevmats.2016.8,10.1073/pnas.2112604118}, while energy injection enables motility and emergent properties of communications and control. For example, colloidal microswimmers direct autonomously in flows  \cite{10.1126/sciadv.1400214, 10.1126/sciadv.aao1755}, leverage machine-learning strategies to navigate through noisy and unexplored environments \cite{10.1126/scirobotics.abd9285} or communicate via clouds of chemical \cite{10.1021/acsnano.9b08421}. Remarkably, proofs of concepts of applications have also already been achieved, highlighting the relevance of the pathway for biomedical applications: swarms micropropellers penetrate the vitreous body of the eye  \cite{10.1126/sciadv.aat4388}, multifunctional rollers deliver cargos in physiological blood flow \cite{10.1126/scirobotics.aba5726} and enzymatic nanomotors can travel {\it in vivo} within the bladder \cite{10.1126/scirobotics.abd2823}. Further examples can be found in recent reviews of the impact of robotics for biomedical applications  \cite{10.1126/scirobotics.abf1462} or therapeutic treatments \cite{10.1038/s41467-020-19322-7,10.1007/s40820-019-0350-5}. Those breakthroughs remain however limited to behavior at the single particle level and lack elementary mechanisms between elements, such as the interlocking of cogwheels. Previous works on cogwheels considered the realization of microfabricated devices powered by active colloids or bacteria  \cite{DiLeonardo:2010kk, Sokolov:2010kt, Maggi:2015dx}, momentum transfer from light\cite{10.1038/s41467-018-06947-y} or light-driven dielectrophoretic forces \cite{10.1038/s41467-021-25582-8, Zhang:2019ei}. They constituted large ($>30\mu m$), monolithic structures, whose direction of rotation was encoded by a rigid template. We previously demonstrated the self-assembly of self-spinning microgears notably limited to a single number of teeth and lacking control on the direction of the rotation \cite{Aubret:2018cca}.



  In contrast, we recently proposed a novel approach that uses templates of optical traps as algorithm to program metamachines, or machines made of machines  \cite{10.1038/s41467-021-26699-6}. In the present work, we leverage this versatile tool to assemble a family of autonomous cogwheels with varying number of teeth and control their chirality using optical vortices of light. We study this family of cogwheel and the behavior of interlocking pairs in contact. They exhibit a diffusive behavior when in contact, the two cogwheels rolling over each other with a mobility, which depends on the curvature. It makes possible for cogwheels to self-position, when interlocked with a structure of variable curvature, which we demonstrate. The ability of the cogwheels to self-position position highlights untapped opportunities for structures and microbots made from active components.  \\

\section*{Results}
\subsection*{Assembly of a Family of Active Cogwheels}
A family of self-spinning cogwheels is assembled using templates of optical traps as algorithms to program machines made of self-propelled colloids \cite{10.1038/s41467-021-26699-6}. In brief, a colloidal bead is optically trapped and decorated with N peripheral optical traps, N being the number of teeth for the cogwheel [Fig.1A]. Active heterodimers travel along the substrate and are captured when crossing the optical traps. The trapping itself is a result of the interplay between alignment by repulsive scattering forces and propulsion, we previously discussed in \cite{10.1038/s41467-021-26699-6}. As a result, we can assemble cogwheels made of a central  sphere surrounded by  $N$ peripheral heterodimers:  the teeth of the cogwheel [Fig.1B, movie S1]. The optical traps are subsequently removed and the cogwheel spins autonomously. The procedure is repeated with different N to form a collection of cogwheels with varying teeth numbers. Practically, we use identical active heterodimers of radii $d$ and vary the radius of the central particle to control the number of teeth $N$ of the cogwheel.  Repeating the procedure for central beads of increasing radii, we find that stable cogwheels correspond to active heterodimers  packed on the periphery of the central sphere, {\it i.e.} $2\pi R=2Nd$. Slight deviations from compact packing yields to unstable structures that disassemble [Fig.1C].  We obtain from timelapse videos, the average rotation rate $\bar\Omega_N$ for self-spinning cogwheels from N=6 to 13 [Fig.1D] and orientation $\theta_N$ of the heterodimers forming the teeth of the cogwheel  [Fig.1E].  Remarkably, we observe that the mean orientation $\theta_N$  follows  $\cos\theta_N\propto 1/R$ (Eq.1) [Fig.1F], highlighting a coupling between orientation of the heterodimers of the cogwheel and curvature. A quantitative understanding of this scaling will require the detailed modeling of hydrodynamics and phoretic interactions in this multi-body systems and is beyond the scope of this paper.  We however use this empirical relationship to relate rotation rate of the cogwheels with their curvature. As the rotation of the cogwheel results from the translational velocity $V_0$ of the active heterodimers, we expect $R\bar\Omega_N\propto V_0 \cos\theta_N$, which gives $\Omega_N\propto 1/R^2$ [Eq.2],  in reasonable agreement with experimental data  [Fig.1D], notwithstanding that the simple geometric argument neglects interactions between active heterodimers. 
\\
\subsection*{Rotation Control by Optical Vortices}
At this point, we demonstrated the assembly of a family of spinning cogwheels, with tunable gears number. After release of the template, the active heterodimers, that  form the teeth, momentarily fluctuate, before a cogwheel starts spinning. The direction of rotation is random,  resulting from a spontaneous symmetry-breaking. Fluctuations in the orientation of the heterodimers can stop the rotation, before reversing direction or proceeding to the same  [movie S2]. The random reversal  of directions is more frequently observed for gears with $N>8$, that exhibit larger fluctuations of orientations  [Fig.1D]. This increase of fluctuations notably originates from lower rotation speeds for cogwheels with larger $N$, in line with experimental results that show a stabilizing effect of hydrodynamics in the orientation of the heterodimers \JP{[FigSxx]}. Our aim is now to control the direction of rotation and to this end we leverage again the versatility of light to manipulate matter at small scale. We use an optical vortex: a helical beam that displays a circular motion with an amount of orbital angular momentum $\ell \hbar$ per photon, where $\ell$ is an integer number and order of the vortex and $\hbar$ is the Planck constant \cite{10.1038/nature01935, 10.1103/physrevlett.90.133901}. The optical vortex induces a  torque on a cogwheel [Fig.2A], transferring orbital angular momentum to the hematite of the heterodimer  \cite{10.1038/s41467-021-26699-6}.  We quantify this torque by measuring the rotation rate of a spinning cogwheel (N=6) illuminated by an optical vortex of opposite chirality and  variable amplitude. Increasing the power of the optical vortex does not affect the geometric arrangement of the  cogwheel, and results in a linear reduction of the angular speed of the cogwheel, as expected from the transfer of angular momentum by the photons [Fig.2B]. The cogwheels notably stalls, $\Omega_6=0\ rad/s$, for an incident power of $P\sim 5mW$, which results in an optical torque exerted by the vortex, $T^*_6 = 6 \alpha_\ell Q_h P/c\sim 6pN.\mu m$, where the $6$ prefactor  accounts for the angular momentum transfer to $N=6$ heterodimers,  $\alpha_\ell$ is a geometric factor accounting for the distribution of light for an optical vortex of order $\ell$ and $Q_h$ is the absorption cross section of the heterodimer and $c$ is the speed of light \JP{[see SI for details on the estimate of $Q_h$, extension of the optical vortex as a function of $\ell$ and derivation]}. This value for the torque from light to stop the rotation is in reasonable agreement with the estimate of the mechanical torque $T\sim 8\pi \eta R^3\Omega_6\sim 8pN.\mu m$ for on a cogwheel of radius $R=4\mu m$ rotating at $\Omega_6\sim 5\ rad/s$ in a fluid of viscosity $\eta=10^{-3} Pas$.
We leverage this optical torque  to set the chirality of the cogwheels and control their direction of rotation. Following assembly or the immobilization of a cogwheel by optical traps [Fig.2C],  the short application  of an  optical vortex ($\sim 100ms$) leads to the orientation of the teeth and sets the direction of rotation of the cogwheel, persisting after removal of the optical vortex [Fig.2C]. The procedure is repeated using optical vortices of opposite helicity  to reverse directions of rotation, exhibiting remarkable reproducibility on the rotation rate  over multiple cycles of direction reversal [Fig.2D, movie S3]. The procedure is similarly extended to cogwheels with different number of teeth and radii by adapting the order of the beam, {\it i.e.} radius of the optical vortex,  to the size of the cogwheel \JP{[SI]}.

\subsection*{Dynamics of Interlocked Cogwheels}

When two microgears are distant, they interact {\it via} phoresis, leading to synchronization mediated by clouds of chemicals. The description of the synchronization can be found in \cite{Aubret:2018cca,Aubret:2018dq}, with the notable predictions that the radial component of the interaction decays rapidly with distance, as $1/r^{N+2}$, where $r$ is the center to center distance between two cogwheels and $N$ is the teeth number.  Distant synchronization furthermore requires to prevent the cogwheels to drift apart by phoretic repulsion   \cite{Aubret:2018cca,Aubret:2018dq}. We instead turn to the study of the  interlocking behavior of pairs of cogwheels in contact and show that it forms a robust and autonomous  mechanism.


We consider a cogwheel with fixed number of teeth,  $N_1=8$,  in contact with a second cogwheel, for which $N_2$ varies  between $N_2=6$ and $N_2=\infty$, the flat interface of a rod-like structure. Using optical traps to manipulate the position of the center of the structures, the cogwheels are set in contact and released from optical traps.  Heterodimers near contact reorient vertically, resulting in an attractive interaction from hydrodynamic pumping \cite{10.1038/s41467-021-26699-6, DiLeonardo:ul, Weinert:2008cn}:  cogwheels of the pair remain in contact and interlock [Fig.3A].  The structure is stable and interlocking cogwheels remain in contact over periods of time $>600s$.  Cogwheels spin and roll over each other, akin to macroscopic cogwheels with interpenetrating gears. The dynamics is however subject to fluctuations: motion stops, and cogwheels can reverse directions [Fig.3B, movie S4]. In order to quantify the motion dynamics, we track the motion of pairs of gears and represent the time evolution of the Mean Squared Angular Displacement (MSAD) $\Delta\phi^2(t)$ for the angle $\phi$ [Fig.3B-inset]. We measure $\Delta\phi^2(t)=2D_N t$ [Fig.3C], indicative of  cogwheels rolling over each other in a random walk with a diffusivity $D_N$ that depends on the number of teeth $N$. The mobility $D_N$ is rapidly decreasing with the number of teeth of the cogwheel or equivalently its curvature, $1/R_N$, and effectively vanish on a flat interface, $R=\infty$ [Fig.3C-inset]. For a cogwheel with rotation rate $\Omega_N$, that is interlocked with an adjacent surface and can reverse direction randomly, we expect to observe a linear MSAD with diffusivity $D_N\propto \Omega^2_N$. Using [Eq.2], $\Omega_N\propto 1/R^2$, we predict a curvature dependent mobility $D_N\propto 1/R^4$, in agreement with our experimental data [Fig.3C-inset].\\

\section*{Discussion}
It results, that a cogwheel interlocked with a structure with variable curvature constitutes  will remain in contact and form a stable pair with a dynamical structure. In order to test this prediction, we study the dynamics of a cogwheel, with N=8, put in contact with a structure with variable curvature [Fig.3E]. The cogwheel travels along the surface of the structure before locking position, where curvature vanishes, {\it i.e.} on the flat portion of the structure. We repeat the experiment to explore configurations and represent the escape rate $1/\tau$, inverse of the time $\tau(X,Y)$ spent  at a given position $(X,Y)$ [Fig.3D].  We successfully reproduce the observed dynamics using a simple model of a random walker with the mobility $D\propto 1/R^4$ measured experimentally on a simplified model of structure with variable curvature [Fig.3E,F, \JP{SI}]. The experiment and model notably predict the self-positioning of active cogwheel on  the flat edge of structures. We leverage this effect to program different microbots from cogwheels and shapes  [Fig.4A, movie S5]. In a first experiment, we assemble two ovoid shapes that are then put in contact with their flat edge with a cogwheel as third structure. Reconfigurations of the formed microbot are absent, the structure sets at contact as a result of the trapping of the cogwheel on flat edges. In contrast, when two cogwheels are put in contact with a third, they roll on the surface until they meet and merge. It results in a different structure from previously, with a reproducible dynamics and final structure. Irrespective of the starting point of the two cogwheels, the parts of the microbots reposition until they form the targeted microbot  [Fig.4A, movie S5].\\ 

We finally exhibit the practical relevance of the achieved microbots for manipulation at small scale. We demonstrate their use to  capture, displace and release a load constituted of a passive sphere [Fig.4B, movie S6]. The load is first encapsulated by a layer of active heterodimers, constituting the central particle of an active cogwheel [Fig.4B]. It  isolates the load from its surrounding and assemble a cogwheel. A structure with variable curvature is assembled separately. The two structures are approached and merged, forming a microbot as discussed above. The microbot exhibits translational motion thanks to the activity of the composing heterodimers \cite{10.1038/s41467-021-26699-6}. The structure autonomously navigates until the load needs to be released [Fig.4C]. It is simply achieved by turning off the activity of the active heterodimers so that the system returns to equilibrium. Particles composing the microbots become passive and diffuse away, releasing the load [Fig.4B, movie S6]. \\

This work demonstrates the potential of active colloids to form autonomous structures, and mechanisms. It shows autonomous machinery with external control: direction of rotation set by optical vortices or migration along optical tracks formed by a light gradient. Using self-positioning components, it unveils an untapped approach to manufacture complex microbots. It highlights how modern robotics can benefit from fundamental advances in colloidal science and active matter.  The ability to direct the mechanism using external stimuli, such as light patterns, or concentrations gradients to achieve cyclic transformations or programmable morphing of structures remains to be explored. Similarly the communications between structures and coordinated movement between parts is a challenge that will need to be addressed towards advanced machineries.  The potential of  active cogwheels to power passive microstructures is a possible avenue of progress towards active microfluidics and robotics at small scale that will similarly benefit from this work.

\newpage
\section*{Materials and Methods}
\subsection*{Materials}
\subsubsection*{Synthesis of hematite cubes}
Synthesis of hematite cubes follows the method described by Sugimoto \cite{Sugimoto:1993bf}. Briefly, we mix 100 mL of 2M FeCl$_3$ $\cdot$ 6H$_2$O, 90 ml 6 NaOH and 10 ml water, in a 250 mL pyrex bottle and shake thoroughly. The bottle is then placed in an oven at $100^\circ$C for 3 to 4 days, until the hematite particles reach desired size. The resulting hematite cubes in the gel network are isolated by successive sedimentation and resuspended in DI water.

\subsubsection*{Synthesis of heterodimers}
Synthesis of heterodimer particles is performed by heterogeneous nucleation of trialkoxysilanes (oil precursor) on hematite particles as seeds. The synthesis procedure is adapted from ref. \cite{Youssef:2016kb}, with chemical modification to reinforce the stability of the heterodimer under light illumination. In particular, we make use of a hydrophobic copolymer Hexadecyltrimethoxysilane (HTS) to chemically protect the bond between the hematite and polymer core against highly reactive hydroxil radicals generated during H$_2$O$_2$ consumption. A beaker with 100 mL of DI water is prepared, and mixed with 120 $\mu$L of a 50\% NH$_3$ solution, giving a pH $\sim 10.5$, under mild magnetic stirring. We add $\sim 1$ mL of an aqueous solution of hematite particles, to get a slightly red-colored solution. Following, we add 100 $\mu$L of HTS,  followed by 500 $\mu$L of 3-(Trimethoxysilyl)propyl methacrylate (TPM). The solution is  covered and stirred for $\sim$ 2h00. During this time, the HTS and TPM hydrolyse, and condense on the hematite particles, with strong wetting leading to their engulfment in the oil phase. After $\sim$ 2h00, the solution is turbid, as a result of the scattering of newly formed colloidal particles. We then add 2 mL of Pluronic F-108 solution (5\% wt), and wait 2 min. Dewetting and extrusion of TPM from the hematite is performed by pH change to $\sim$ 2.1, by adding 1.5 mL of 1M chloridric acid HCl. The solution is let under stirring for 3 mn, and diluted 4 times. We then carry out the polymerization by adding 50 mg of radical initiator Azobisisobutyronitrile (AIBN) to the solution and leave it under stirring for $\sim 5$ mn. The beaker is covered with an aluminium foil and placed in a pre-heated oven at $\sim$ 80 degrees Celsius for 2 hours. We let the solution cool down to room temperature, remove the excess solution above the sedimented particles, and add 50 mL DI water with 1 mL of 250 mM NaOH solution, giving a pH $\sim$ 10, and let the solution overnight to facilitate the hydrolysis and condensation of any remaining HTS monomers. The solution is then centrifuged and rinsed multiple times to remove the excess TPM/HTS particles and obtain the desired colloidal solution of heterodimers. 

\subsection*{Materials}
 \subsubsection*{Sample preparation}
The heterodimer particles are diluted in a 6\% solution of hydrogen peroxide H$_2$O$_2$ (Fisher Scientific H325-500) in deionized water (Milli-Q, resistivity 18.2M).  The samples are prepared at low particle density of $\sim 10^{-3}$ $\mu$m$^{-2}$.  The cell for the solution is composed of a rectangular capillary glass (VitroCom 3520-050), previously plasma cleaned (Harrick Plasma PDC-001) and thoroughly rinsed with DI water. The solution is injected in the capillary, then sealed with capillary wax (Hampton Research HR4-328). Particles sediment near the bottom surface of the capillary, and observations with the microscope are made in this plane.

\subsubsection*{Optical setup}
 The experiments are carried out on a custom-made optical setup, allowing for simultaneous uniform excitation of the microswimmers and holographic optical tweezing. The sample is observed with bright-field transmitted illumination (LED1). A LED is set up in the blue range (LED2, $\lambda = 425 - 500
$ nm, Lumencor SOLA 6-LCR-SC) and uniformly illuminates the sample on a large area (size $\sim 300 \times 300$ $\mu$m$^2$) to activate the swimmers by triggering the photocatalytic decomposition of H$_2$O$_2$ by the hematite (typical intensity of $\sim 1 \, \mu$W/$\mu$m$^2$). A red continuous laser with near TEM00 mode ($\lambda = 639$ nm, Coherent, Genesis MX639-1000 STM, $M^2 < 1.1$) is added on the optical path. The linearly polarized beam is collimated, rotated with a $\lambda/2$ waveplate, and sent on the surface of a Spatial Light Modulator (SLM, Holoeye -LETO Phase Only Spatial Light Modulator). The optical path follows a typical 4-f setup using two $f_1=400$ mm lenses, and the zero-order of the diffracted beam is filtered out with a diaphragm at equal distance $d=f$ between the two lenses. Following, a hologram is formed at the back aperture of a high NA objective (Nikon apo-TIRF, $\times 100$, NA=1.45) allowing for the creation of complex spatiotemporal optical patterns in the object plane, at the bottom surface of the sample. The hologram is computed in real time using a computer software (Holoeye), with the phase patterns computed under Matlab, and allows for the selective trapping and manipulation of individual swimmers. 
An electronic shutter (Thorlabs SHB1T) on the red optical path enables switching ON and OFF the laser traps.
 The sample is mounted on a manual micrometric stage (Nikon Ti-SR). Observation is performed through the same objective as for excitation, and the bright-field signal is reflected on a polarizing beam splitter and observed on 2 monochrome Charged Coupled Devices (CCDs) with different resolutions (0.05 $\mu$m/px and 0.1 $\mu$m/px, respectively, Edmund Optics EO-1312M), with appropriate spectral filters.\JP{Adding optical vortex part}. \JP{Scheme of the optical setup is presented in SI.}

\subsubsection*{Image acquisition and analysis}
 All experiments are recorded on the CCD camera with 0.05 $\mu$m/px resolution, at frame rate between 20-50 fps, and under bright field illumination. When necessary, tracking was performed separately for the hematite and TPM particles with custom-made Matlab routine.
Error bars are obtained from standard deviations of experimental measurements.

\subsection*{Data availability}
The data that support the plots within this paper and other findings of this study are available from the corresponding authors upon request.

\subsection*{Code availability}
The code that support the plots within this paper and other findings of this study are available from the corresponding authors upon request.

\newpage

\bibliographystyle{naturemag}

\bibliography{bibJP.bib}

\begin{thebibliography}{10}
\expandafter\ifx\csname url\endcsname\relax
  \def\url#1{\texttt{#1}}\fi
\expandafter\ifx\csname urlprefix\endcsname\relax\def\urlprefix{URL }\fi
\providecommand{\bibinfo}[2]{#2}
\providecommand{\eprint}[2][]{\url{#2}}

\bibitem{Houdusse_nanomotors}
\bibinfo{author}{Houdusse, A.}
\newblock \bibinfo{title}{{Biological nanomotors, driving forces of life}}.
\newblock \emph{\bibinfo{journal}{Comptes Rendus. Biologies}}
  \textbf{\bibinfo{volume}{343}}, \bibinfo{pages}{53--78}
  (\bibinfo{year}{2021}).

\bibitem{chromosomes_loops}
\bibinfo{author}{Fudenberg, G.} \emph{et~al.}
\newblock \bibinfo{title}{{Formation of Chromosomal Domains by Loop
  Extrusion}}.
\newblock \emph{\bibinfo{journal}{Cell Reports}} \textbf{\bibinfo{volume}{15}},
  \bibinfo{pages}{2038--2049} (\bibinfo{year}{2016}).

\bibitem{Alberts:887198}
\bibinfo{author}{Alberts, B.}, \bibinfo{author}{Bray, D.} \&
  \bibinfo{author}{Lewis, J.}
\newblock \emph{\bibinfo{title}{{Molecular biology of the cell; 2nd ed.}}}
\newblock Garland (\bibinfo{publisher}{Garland}, \bibinfo{year}{1989}).

\bibitem{Li:2019fk}
\bibinfo{author}{Li, S.} \emph{et~al.}
\newblock \bibinfo{title}{{Particle robotics based on statistical mechanics of
  loosely coupled components}}.
\newblock \emph{\bibinfo{journal}{Nature}} \textbf{\bibinfo{volume}{567}},
  \bibinfo{pages}{1 -- 6} (\bibinfo{year}{2019}).

\bibitem{Rubenstein:2014dt}
\bibinfo{author}{Rubenstein, M.}, \bibinfo{author}{Cornejo, A.} \&
  \bibinfo{author}{Nagpal, R.}
\newblock \bibinfo{title}{{Robotics. Programmable self-assembly in a
  thousand-robot swarm.}}
\newblock \emph{\bibinfo{journal}{Science}} \textbf{\bibinfo{volume}{345}},
  \bibinfo{pages}{795 -- 799} (\bibinfo{year}{2014}).

\bibitem{10.1021/jacs.1c04836}
\bibinfo{author}{Ye, Z.} \emph{et~al.}
\newblock \bibinfo{title}{{Construction of Nanomotors with Replaceable Engines
  by Supramolecular Machine-Based Host-Guest Assembly and Disassembly.}}
\newblock \emph{\bibinfo{journal}{Journal of the American Chemical Society}}
  \textbf{\bibinfo{volume}{143}}, \bibinfo{pages}{15063--15072}
  (\bibinfo{year}{2021}).

\bibitem{10.1002/adma.202101965}
\bibinfo{author}{Joh, H.} \& \bibinfo{author}{Fan, D.~E.}
\newblock \bibinfo{title}{{Materials and Schemes of Multimodal Reconfigurable
  Micro/Nanomachines and Robots: Review and Perspective.}}
\newblock \emph{\bibinfo{journal}{Advanced materials}}
  \textbf{\bibinfo{volume}{33}}, \bibinfo{pages}{e2101965}
  (\bibinfo{year}{2021}).

\bibitem{He:2020kn}
\bibinfo{author}{He, M.} \emph{et~al.}
\newblock \bibinfo{title}{{Colloidal diamond.}}
\newblock \emph{\bibinfo{journal}{Nature}} \textbf{\bibinfo{volume}{585}},
  \bibinfo{pages}{524 -- 529} (\bibinfo{year}{2020}).

\bibitem{10.1038/s41578-021-00323-x}
\bibinfo{author}{Hueckel, T.}, \bibinfo{author}{Hocky, G.~M.} \&
  \bibinfo{author}{Sacanna, S.}
\newblock \bibinfo{title}{{Total synthesis of colloidal matter}}.
\newblock \emph{\bibinfo{journal}{Nature Reviews Materials}}
  \textbf{\bibinfo{volume}{6}}, \bibinfo{pages}{1053--1069}
  (\bibinfo{year}{2021}).

\bibitem{Chaudhary:2012et}
\bibinfo{author}{Chaudhary, K.}, \bibinfo{author}{Chen, Q.},
  \bibinfo{author}{Juarez, J.~J.}, \bibinfo{author}{Granick, S.} \&
  \bibinfo{author}{Lewis, J.~A.}
\newblock \bibinfo{title}{{Janus Colloidal Matchsticks}}.
\newblock \emph{\bibinfo{journal}{Journal of the American Chemical Society}}
  \textbf{\bibinfo{volume}{134}}, \bibinfo{pages}{12901 -- 12903}
  (\bibinfo{year}{2012}).

\bibitem{10.1038/natrevmats.2016.8}
\bibinfo{author}{Rogers, W.~B.}, \bibinfo{author}{Shih, W.~M.} \&
  \bibinfo{author}{Manoharan, V.~N.}
\newblock \bibinfo{title}{{Using DNA to program the self-assembly of colloidal
  nanoparticles and microparticles}}.
\newblock \emph{\bibinfo{journal}{Nature Reviews Materials}}
  \textbf{\bibinfo{volume}{1}}, \bibinfo{pages}{16008} (\bibinfo{year}{2016}).

\bibitem{10.1073/pnas.2112604118}
\bibinfo{author}{McMullen, A.}, \bibinfo{author}{Hilgenfeldt, S.} \&
  \bibinfo{author}{Brujic, J.}
\newblock \bibinfo{title}{{DNA self-organization controls valence in
  programmable colloid design}}.
\newblock \emph{\bibinfo{journal}{Proceedings of the National Academy of
  Sciences}} \textbf{\bibinfo{volume}{118}}, \bibinfo{pages}{e2112604118}
  (\bibinfo{year}{2021}).

\bibitem{10.1126/sciadv.1400214}
\bibinfo{author}{Palacci, J.} \emph{et~al.}
\newblock \bibinfo{title}{{Artificial rheotaxis}}.
\newblock \emph{\bibinfo{journal}{Science Advances}}
  \textbf{\bibinfo{volume}{1}}, \bibinfo{pages}{e1400214 -- e1400214}
  (\bibinfo{year}{2015}).

\bibitem{10.1126/sciadv.aao1755}
\bibinfo{author}{Katuri, J.}, \bibinfo{author}{Uspal, W.~E.},
  \bibinfo{author}{Simmchen, J.}, \bibinfo{author}{Miguel-Lopez, A.} \&
  \bibinfo{author}{Sanchez, S.}
\newblock \bibinfo{title}{{Cross-stream migration of active particles}}.
\newblock \emph{\bibinfo{journal}{Science Advances}}
  \textbf{\bibinfo{volume}{4}}, \bibinfo{pages}{eaao1755}
  (\bibinfo{year}{2018}).
\newblock \eprint{1706.06817}.

\bibitem{10.1126/scirobotics.abd9285}
\bibinfo{author}{Muinos-Landin, S.}, \bibinfo{author}{Fischer, A.},
  \bibinfo{author}{Holubec, V.} \& \bibinfo{author}{Cichos, F.}
\newblock \bibinfo{title}{{Reinforcement learning with artificial
  microswimmers}}.
\newblock \emph{\bibinfo{journal}{Science Robotics}}
  \textbf{\bibinfo{volume}{6}}, \bibinfo{pages}{eabd9285}
  (\bibinfo{year}{2021}).

\bibitem{10.1021/acsnano.9b08421}
\bibinfo{author}{Zhou, C.} \emph{et~al.}
\newblock \bibinfo{title}{{Coordinating an Ensemble of Chemical Micromotors via
  Spontaneous Synchronization}}.
\newblock \emph{\bibinfo{journal}{ACS Nano}} \textbf{\bibinfo{volume}{14}},
  \bibinfo{pages}{5360--5370} (\bibinfo{year}{2020}).

\bibitem{10.1126/sciadv.aat4388}
\bibinfo{author}{Wu, Z.} \emph{et~al.}
\newblock \bibinfo{title}{{A swarm of slippery micropropellers penetrates the
  vitreous body of the eye}}.
\newblock \emph{\bibinfo{journal}{Science Advances}}
  \textbf{\bibinfo{volume}{4}}, \bibinfo{pages}{eaat4388}
  (\bibinfo{year}{2018}).

\bibitem{10.1126/scirobotics.aba5726}
\bibinfo{author}{Alapan, Y.}, \bibinfo{author}{Bozuyuk, U.},
  \bibinfo{author}{Erkoc, P.}, \bibinfo{author}{Karacakol, A.~C.} \&
  \bibinfo{author}{Sitti, M.}
\newblock \bibinfo{title}{{Multifunctional surface microrollers for targeted
  cargo delivery in physiological blood flow}}.
\newblock \emph{\bibinfo{journal}{Science Robotics}}
  \textbf{\bibinfo{volume}{5}}, \bibinfo{pages}{eaba5726}
  (\bibinfo{year}{2020}).

\bibitem{10.1126/scirobotics.abd2823}
\bibinfo{author}{Hortelao, A.~C.} \emph{et~al.}
\newblock \bibinfo{title}{{Swarming behavior and in vivo monitoring of
  enzymatic nanomotors within the bladder.}}
\newblock \emph{\bibinfo{journal}{Science robotics}}
  \textbf{\bibinfo{volume}{6}} (\bibinfo{year}{2021}).

\bibitem{10.1126/scirobotics.abf1462}
\bibinfo{author}{Gao, A.} \emph{et~al.}
\newblock \bibinfo{title}{{Progress in robotics for combating infectious
  diseases}}.
\newblock \emph{\bibinfo{journal}{Science Robotics}}
  \textbf{\bibinfo{volume}{6}}, \bibinfo{pages}{eabf1462}
  (\bibinfo{year}{2021}).

\bibitem{10.1038/s41467-020-19322-7}
\bibinfo{author}{Schmidt, C.~K.}, \bibinfo{author}{Medina-Sanchez, M.},
  \bibinfo{author}{Edmondson, R.~J.} \& \bibinfo{author}{Schmidt, O.~G.}
\newblock \bibinfo{title}{{Engineering microrobots for targeted cancer
  therapies from a medical perspective}}.
\newblock \emph{\bibinfo{journal}{Nature Communications}}
  \textbf{\bibinfo{volume}{11}}, \bibinfo{pages}{5618} (\bibinfo{year}{2020}).

\bibitem{10.1007/s40820-019-0350-5}
\bibinfo{author}{Wang, J.}, \bibinfo{author}{Dong, R.}, \bibinfo{author}{Wu,
  H.}, \bibinfo{author}{Cai, Y.} \& \bibinfo{author}{Ren, B.}
\newblock \bibinfo{title}{{A Review on Artificial Micro/Nanomotors for
  Cancer-Targeted Delivery, Diagnosis, and Therapy}}.
\newblock \emph{\bibinfo{journal}{Nano-Micro Letters}}
  \textbf{\bibinfo{volume}{12}}, \bibinfo{pages}{11} (\bibinfo{year}{2020}).

\bibitem{DiLeonardo:2010kk}
\bibinfo{author}{Leonardo, R.~D.} \emph{et~al.}
\newblock \bibinfo{title}{{Bacterial ratchet motors}}.
\newblock \emph{\bibinfo{journal}{Proceedings of the National Academy of
  Sciences of the U.S.A}} \textbf{\bibinfo{volume}{107}}, \bibinfo{pages}{9541
  -- 9545} (\bibinfo{year}{2010}).

\bibitem{Sokolov:2010kt}
\bibinfo{author}{Sokolov, A.}, \bibinfo{author}{Apodaca, M.~M.},
  \bibinfo{author}{Grzybowski, B.~A.} \& \bibinfo{author}{Aranson, I.~S.}
\newblock \bibinfo{title}{{Swimming bacteria power microscopic gears}}.
\newblock \emph{\bibinfo{journal}{Proceedings of the National Academy of
  Sciences of the U.S.A}} \textbf{\bibinfo{volume}{107}}, \bibinfo{pages}{969
  -- 974} (\bibinfo{year}{2010}).

\bibitem{Maggi:2015dx}
\bibinfo{author}{Maggi, C.} \emph{et~al.}
\newblock \bibinfo{title}{{Self-Assembly of Micromachining Systems Powered by
  Janus Micromotors}}.
\newblock \emph{\bibinfo{journal}{Small}} \textbf{\bibinfo{volume}{12}},
  \bibinfo{pages}{446 -- 451} (\bibinfo{year}{2015}).

\bibitem{10.1038/s41467-018-06947-y}
\bibinfo{author}{Bianchi, S.}, \bibinfo{author}{Vizsnyiczai, G.},
  \bibinfo{author}{Ferretti, S.}, \bibinfo{author}{Maggi, C.} \&
  \bibinfo{author}{Leonardo, R.~D.}
\newblock \bibinfo{title}{{An optical reaction micro-turbine.}}
\newblock \emph{\bibinfo{journal}{Nature communications}}
  \textbf{\bibinfo{volume}{9}}, \bibinfo{pages}{4476} (\bibinfo{year}{2018}).

\bibitem{10.1038/s41467-021-25582-8}
\bibinfo{author}{Zhang, S.} \emph{et~al.}
\newblock \bibinfo{title}{{Reconfigurable multi-component micromachines driven
  by optoelectronic tweezers}}.
\newblock \emph{\bibinfo{journal}{Nature Communications}}
  \textbf{\bibinfo{volume}{12}}, \bibinfo{pages}{5349} (\bibinfo{year}{2021}).

\bibitem{Zhang:2019ei}
\bibinfo{author}{Zhang, S.} \emph{et~al.}
\newblock \bibinfo{title}{{The optoelectronic microrobot: A versatile toolbox
  for micromanipulation}}.
\newblock \emph{\bibinfo{journal}{Proceedings of the National Academy of
  Sciences}} \textbf{\bibinfo{volume}{116}}, \bibinfo{pages}{14823--14828}
  (\bibinfo{year}{2019}).

\bibitem{Aubret:2018cca}
\bibinfo{author}{Aubret, A.}, \bibinfo{author}{Youssef, M.},
  \bibinfo{author}{Sacanna, S.} \& \bibinfo{author}{Palacci, J.}
\newblock \bibinfo{title}{{Targeted assembly and synchronization of
  self-spinning microgears}}.
\newblock \emph{\bibinfo{journal}{Nature Physics}}
  \textbf{\bibinfo{volume}{14}}, \bibinfo{pages}{1 -- 5}
  (\bibinfo{year}{2018}).

\bibitem{10.1038/s41467-021-26699-6}
\bibinfo{author}{Aubret, A.}, \bibinfo{author}{Martinet, Q.} \&
  \bibinfo{author}{Palacci, J.}
\newblock \bibinfo{title}{{Metamachines of pluripotent colloids}}.
\newblock \emph{\bibinfo{journal}{Nature Communications}}
  \textbf{\bibinfo{volume}{12}}, \bibinfo{pages}{6398} (\bibinfo{year}{2021}).

\bibitem{10.1038/nature01935}
\bibinfo{author}{Grier, D.~G.}
\newblock \bibinfo{title}{{A revolution in optical manipulation}}.
\newblock \emph{\bibinfo{journal}{Nature}} \textbf{\bibinfo{volume}{424}},
  \bibinfo{pages}{810--816} (\bibinfo{year}{2003}).

\bibitem{10.1103/physrevlett.90.133901}
\bibinfo{author}{Curtis, J.~E.} \& \bibinfo{author}{Grier, D.~G.}
\newblock \bibinfo{title}{{Structure of Optical Vortices}}.
\newblock \emph{\bibinfo{journal}{Physical Review Letters}}
  \textbf{\bibinfo{volume}{90}}, \bibinfo{pages}{133901}
  (\bibinfo{year}{2003}).

\bibitem{Aubret:2018dq}
\bibinfo{author}{Aubret, A.} \& \bibinfo{author}{Palacci, J.}
\newblock \bibinfo{title}{{Diffusiophoretic design of self-spinning microgears
  from colloidal microswimmers}}.
\newblock \emph{\bibinfo{journal}{Soft Matter}} \textbf{\bibinfo{volume}{14}},
  \bibinfo{pages}{9577 -- 9588} (\bibinfo{year}{2018}).

\bibitem{DiLeonardo:ul}
\bibinfo{author}{Leonardo, R.~D.}, \bibinfo{author}{Ianni, F.} \&
  \bibinfo{author}{Ruocco, G.}
\newblock \bibinfo{title}{{Colloidal Attraction Induced by a Temperature
  Gradient}}.
\newblock \emph{\bibinfo{journal}{Langmuir}} \textbf{\bibinfo{volume}{25}},
  \bibinfo{pages}{4247 -- 4250} (\bibinfo{year}{2009}).

\bibitem{Weinert:2008cn}
\bibinfo{author}{Weinert, F.~M.} \& \bibinfo{author}{Braun, D.}
\newblock \bibinfo{title}{{Observation of Slip Flow in Thermophoresis}}.
\newblock \emph{\bibinfo{journal}{Physical Review Letters}}
  \textbf{\bibinfo{volume}{101}}, \bibinfo{pages}{168301}
  (\bibinfo{year}{2008}).

\bibitem{Sugimoto:1993bf}
\bibinfo{author}{Sugimoto, T.}, \bibinfo{author}{Sakata, K.} \&
  \bibinfo{author}{Muramatsu, A.}
\newblock \bibinfo{title}{{Formation Mechanism of Monodisperse Pseudocubic
  Fe2O3 Particles from Condensed Ferric Hydroxide Gel}}.
\newblock \emph{\bibinfo{journal}{Journal Of Colloid And Interface Science}}
  \textbf{\bibinfo{volume}{159}}, \bibinfo{pages}{372 -- 382}
  (\bibinfo{year}{1993}).

\bibitem{Youssef:2016kb}
\bibinfo{author}{Youssef, M.}, \bibinfo{author}{Hueckel, T.},
  \bibinfo{author}{Yi, G.-R.} \& \bibinfo{author}{Sacanna, S.}
\newblock \bibinfo{title}{{Shape-shifting colloids via stimulated dewetting}}.
\newblock \emph{\bibinfo{journal}{Nature Communications}}
  \textbf{\bibinfo{volume}{7}}, \bibinfo{pages}{1 -- 7} (\bibinfo{year}{2016}).

\end{thebibliography}

\section*{Acknowledgments}
This material is based upon work supported by the Department of Army Research under grant W911NF-20-1-0112.


\section*{Competing Interests} 
The authors declare that they have no competing financial interests.

\newpage

\begin{figure}
\centering
\includegraphics[scale=.8]{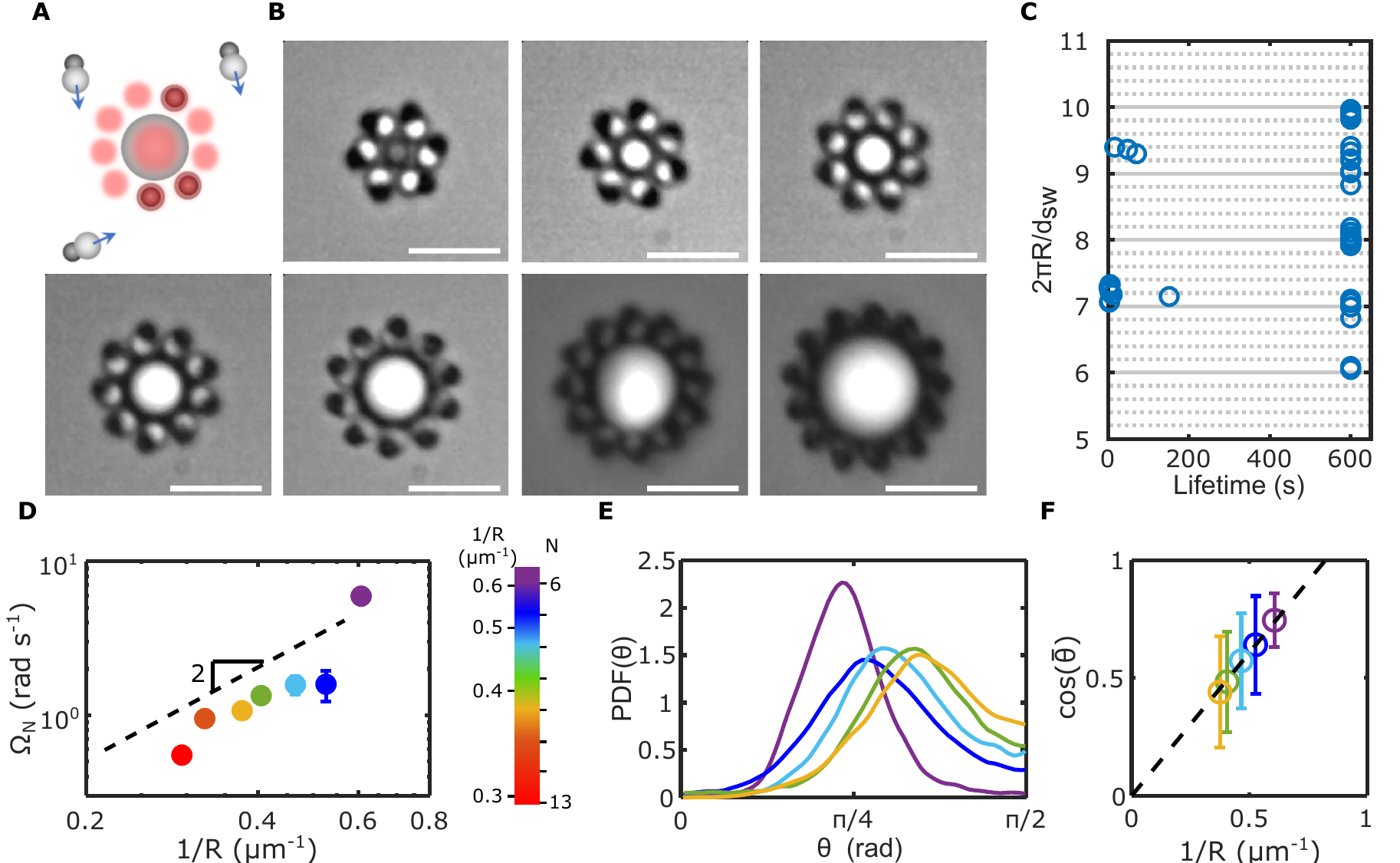}
\label{Fig1}
\caption{\footnotesize\textbf{Family of Spinning Cogwheels}.\textbf{(A)} Formation of a cogwheel using templates of optical traps (in red). A bead is trapped in a center trap and decorated with optical traps at the periphery, then occupied by active heterodimers. Following their occupancy, the traps are removed, and the structure starts spinning, forming an autonomous cogwheel. \textbf{(B)} Family of Cogwheels for N=6 to N=13.  Scale bar is $5\mu m$. \textbf{(C)} Cogwheels which perimeter is an integer of the heterodimers diameter $d$ are stable and are observed over 600s, other structures quickly collapse. \textbf{(D)} Amplitude of the Angular Velocity $\Omega_N$ of the rotation of the cogwheels, for different number of teeth N=6-13. Comparison of the experimental data with predicted scaling $\Omega_N \propto 1/R^2$ [dashed line, see main text]. \textbf{(D)} Probability Distribution Function (PDF) of the orientation $\theta$ of the heterodimers of the cogwheels for variable teeth number $N$. \textbf{(E)} The mean value of the distribution $\bar\theta$ obtained from the PDF scales as $\cos \bar\theta\propto 1/R$.  
}
\end{figure}

\begin{figure}
\centering
\includegraphics[scale=.8]{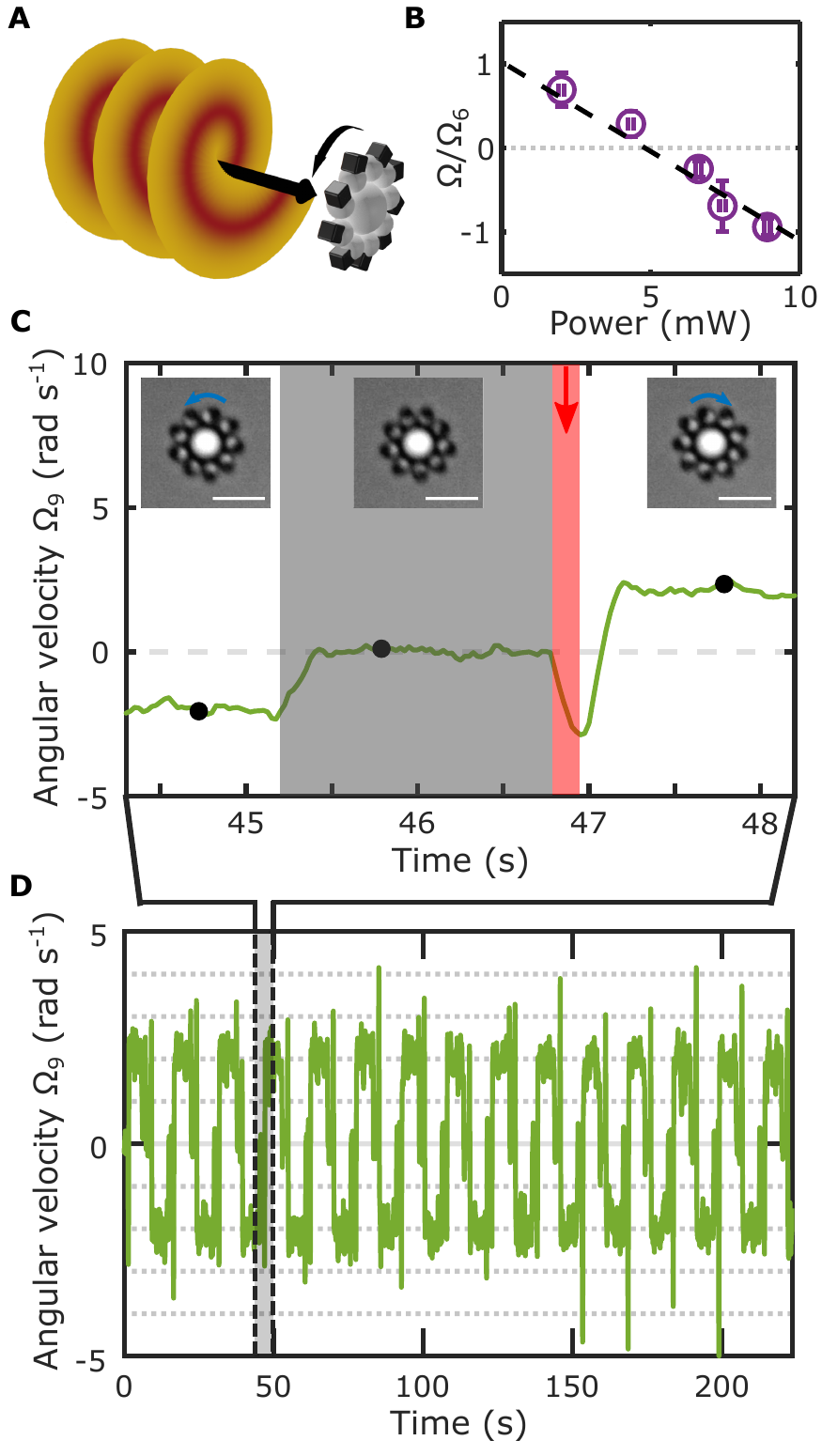}
\label{Fig2}
\caption{\footnotesize\textbf{ Chirality control of Cogwheels using Optical Vortices.} \textbf{(A)} 3D-rendering of a helical beam, {\it i.e.} an optical vortex, and optical momentum transfer to a cogwheel (in grey).   \textbf{(B)} Normalized angular velocity of a cogwheel, with N=6, subject to the optical torque from a superimposed optical vortex of opposite chirality for varying power. The angular velocity of the cogwheel linearly reduces with incident light power, confirming the opposite torque imposed by optical vortex. The angular velocity vanishes for $P\sim 5mW$, when optical torque of the vortex balances propulsion [see main text].  \textbf{(C)} Experimental procedure leading to the control of chirality of the cogwheels using Optical Vortices and measurement of the rotation velocity of a cogwheel, with N=9 (green trace). The cogwheel is initially spinning counter-clockwise rotation. It is shortly  immobilized ($\sim 1s$) by a template of optical traps (grey area). The traps are removed and an optical vortex is applied briefly ($\sim 100ms$, red area and arrow), applying a torque, that reorient the teeth and set the chirality of the cogwheel. The cogwheel spins clockwise. In inset, bright field images of the cogwheel corresponding to the black dot on the timeline. Scale bar is $5\mu m$. \textbf{(D)} The procedure is repeated to control the chirality over 15 cycles of reversal, exhibiting remarkable reproducibility of the rotation velocity. The shaded area corresponds to the timeline shown in (C).}
\end{figure}

\begin{figure}
\centering
\includegraphics[scale=.8]{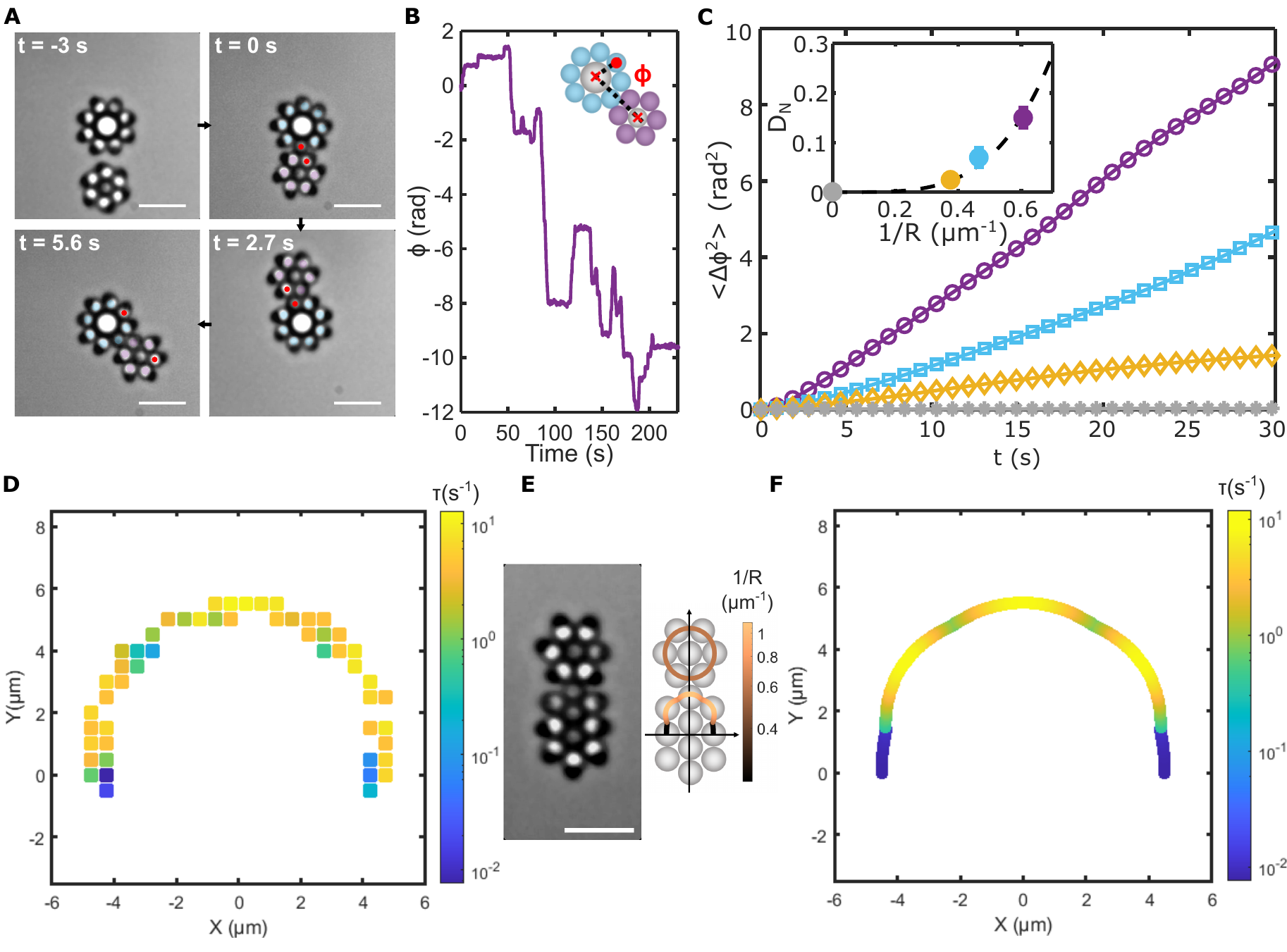}
\label{Fig3}
\caption{\footnotesize\textbf{Interlocking cogwheels in contact.}  \textbf{(A)} Timelapse of interlocking cogwheels. After assembly, they are put in contact using optical traps then released. They interlock and remain in contact, rolling over each other. The red dot is tethered to a teeth and a visual guide to see the rolling.   \textbf{(B-inset)} Definition of the angle $\Phi$ measuring the relative position of the cogwheels.  \textbf{(B)} Time evolution of $\Phi(t)$ measured experimentally, showing the rolling of the cogwheels, as well as reversal of directions of rotation.  \textbf{(C)} Time evolution of the Mean Squared Angular Displacement (MSAD) $\Delta \Phi^2(t)$ for a central cogwheel, with N=8, interlocked with cogwheel of increasing curvature. The MSAD increases linearly with time, indicative of an angular random walk with a curvature dependent mobility $D_N$.
\textbf{(C-inset)} Measurement of Diffusivity $D_N$ from MSAD and comparison to the predicted scaling $D\propto 1/R^4$ [dashed line, see main text].  \textbf{(D-F) Dynamics of a cogwheel interlocked with a structure of variable curvature.} \textbf{(D)} Experimental measurement of $1/\tau$ along the structure of variable curvature, where $\tau$ is the time spent by the cogwheel at position $(X,Y)$.  \textbf{(E-left)} Bright-field image of a cogwheel  interlocked with a structure of variable curvature. Scale bar is $5\mu m$. \textbf{(E-right)} Simplified model of the structure in (E-left) and curvature value (colorbar). \textbf{(F)} Numerical prediction of $1/\tau$, where $\tau$ is the time spent by the cogwheel at position $(X,Y)$, using the measured relationship $D\propto 1/R^4$ on the simplified structure depicted in (E-right). We observe a good agreement with the experimental value. Experiment and numerical results notably exhibits a stable configuration on areas of vanishing curvature.}
\end{figure}

\begin{figure}
\centering
\includegraphics[scale=.8]{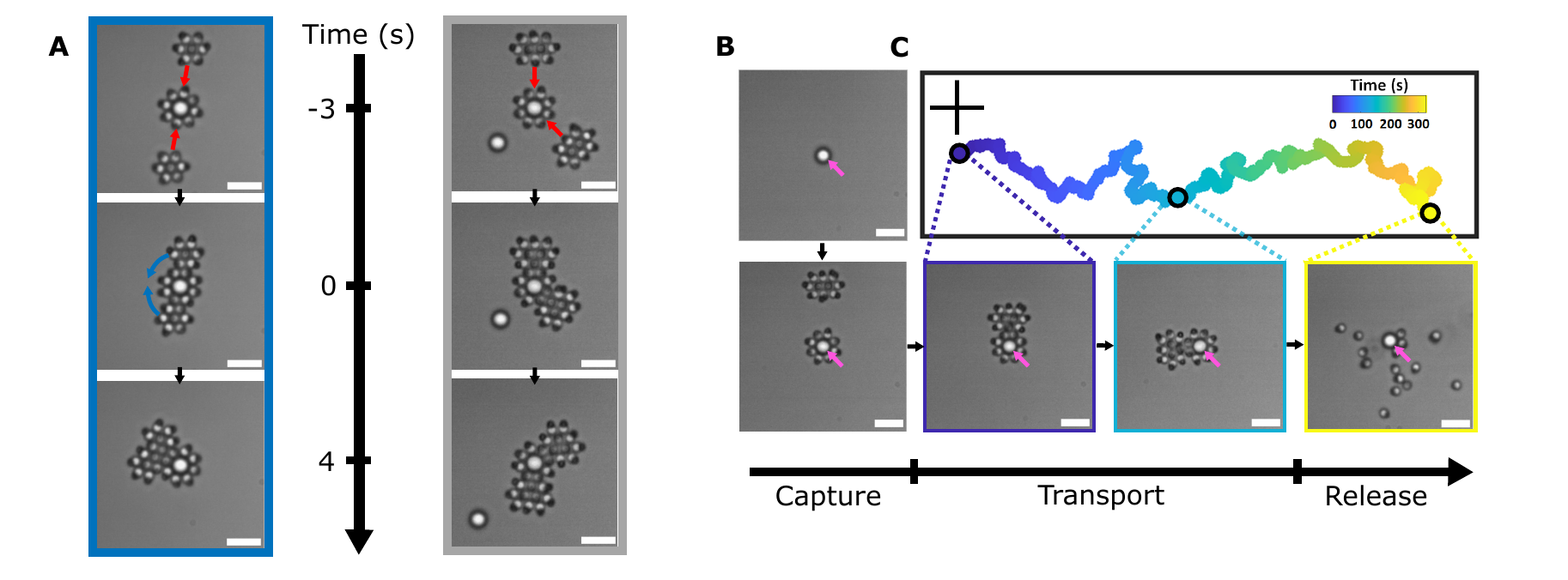}
\label{Fig4}
\caption{\footnotesize\textbf{Reprogrammable Microbots and Cargo Transport. }\textbf{(A)} \textbf{Self-positioning of structures}. A central cogwheel, with N=8, is interlocked with two structures with a flat edge (left) or two cogwheels, with N=6 and constant curvature. (A-left) It forms a stable structure, which shape does not change. (A-right), the cogwheels roll over the central cogwheel and self-position to form a targeted microbot. \textbf{(B,C)} \textbf{Cargo Transport and Release.}  \textbf{(B)} Timelapse bright field images. A passive cargo (pink arrow) is encapsulated and surrounded by active heterodimers to form a cogwheel. The cogwheel is merged to form a microbot with translational motion. The microbot navigates and the load is released on demand by simply turning off the activity of the swimmers. \textbf{(B)}  Trajectory of the microbot transporting the cargo and  bright field images of the system at the time highlighted on the trajectory. Cross is a  $20\mu m$ scale bar. }
\end{figure}

\end{document}